\PassOptionsToPackage{hyphens}{url}
\documentclass[a4paper,USenglish,cleveref, autoref, thm-restate]{lipics-v2021}

\pdfoutput=1 
\hideLIPIcs  

\usepackage[newfloat=true,frozencache=true,cachedir=_minted_cache]{minted} 
\makeatletter
\@ifpackagelater{minted}{2024/09/22}
  { 
    \setminted{bgcolor=lipicsLightGray,bgcolorpadding=\fboxsep,fontsize=\small}
    \captionsetup[listing]{skip=-9pt}  
  }
  { 
    \setminted{bgcolor=lipicsLightGray,fontsize=\small}
    \captionsetup[listing]{skip=-5pt}  
  }
\makeatother

\usepackage{iftex}
\ifluatex
\usepackage{emoji}
\def\YES{\emoji{check-mark-button}}
\def\YESx{\emoji{check-box-with-check}}
\def\NO{\emoji{cross-mark}}
\def\WARN{\emoji{warning}}
\else
\usepackage{bbding}
\definecolor{darkgreen}{rgb}{0.00,0.50,0.00}
\definecolor{darkcyan}{rgb}{0.00,0.20,0.80}
\definecolor{darkyellow}{rgb}{0.99,0.70,0.00}
\def\YES{{\textcolor{darkgreen}\Checkmark}}
\def\YESx{{\textcolor{darkcyan}\CheckmarkBold}}
\def\NO{{\textcolor{red}\XSolidBrush}}
\def\WARN{{\textcolor{darkyellow}\TriangleUp}}
\fi
\usepackage{tablefootnote}

\title{A first look at ROS~2 applications written in asynchronous Rust}

\author{Martin Škoudlil}{Czech Institute of Informatics, Robotics and Cybernetics,\\Czech Technical University in Prague, Czech Republic}{skoudmar@cvut.cz}{}{}

\author{Michal Sojka\footnote{corresponding author}}{Czech Institute of Informatics, Robotics and Cybernetics,\\Czech Technical University in Prague, Czech Republic}{michal.sojka@cvut.cz}{https://orcid.org/0000-0002-8738-075X}{}

\author{Zdeněk Hanzálek}{Czech Institute of Informatics, Robotics and Cybernetics,\\Czech Technical University in Prague, Czech Republic}{zdenek.hanzalek@cvut.cz}{https://orcid.org/0000-0002-8135-1296}{This work was co-funded by the European Union under the project ROBOPROX (reg. no. CZ.02.01.01/00/22\_008/0004590).
}

\authorrunning{M. Škoudlil, M. Sojka and Z. Hanzálek} 

\Copyright{Martin Škoudlil, Michal Sojka and Zdeněk Hanzálek} 



\category{} 

\relatedversion{} 

\supplement{}
\supplementdetails[subcategory={Source code}, cite={}, swhid={}]{Software}{https://zenodo.org/records/15446666}

\funding{This work was supported by the Technology Agency of the Czech Republic under the project Certicar CK03000033.}

\acknowledgements{We want to thank our reviewers and a shepherd for valuable comments that lead to improvement of this paper.}

\nolinenumbers 

\EventEditors{Renato Mancuso}
\EventNoEds{1}
\EventLongTitle{37th Euromicro Conference on Real-Time Systems (ECRTS 2025)}
\EventShortTitle{ECRTS 2025}
\EventAcronym{ECRTS}
\EventYear{2025}
\EventDate{July 8--11, 2025}
\EventLocation{Brussels, Belgium}
\EventLogo{}
\SeriesVolume{335}
\ArticleNo{15}

\begin{document}

\maketitle

\begin{abstract}
  The increasing popularity of the Rust programming language in building robotic applications using the Robot Operating System (ROS~2) raises questions about its real-time execution capabilities, particularly when employing asynchronous programming. Existing real-time scheduling and response-time analysis techniques for ROS~2 focus on applications written in C++ and do not address the unique execution models and challenges presented by Rust's asynchronous programming paradigm. In this paper, we analyze the execution model of R2R -- an asynchronous Rust ROS~2 bindings and various asynchronous Rust runtimes, comparing them with the execution model of C++ ROS~2 applications. We propose a structured approach for R2R applications aimed at deterministic real-time operation involving thread prioritization and callback-to-thread mapping schemes. Our experimental evaluation based on measuring end-to-end latencies of a synthetic application shows that the proposed approach is effective and outperforms other evaluated configurations. A more complex autonomous driving case study demonstrates its practical applicability. Overall, the experimental results indicate that our proposed structure achieves bounded response times for time-critical tasks. This paves the way for future work to adapt existing or develop new response-time analysis techniques for R2R applications using our structure.
\end{abstract}

\section{Introduction}

The Robot Operating System (ROS)~\cite{2024:ROS2Jazzy} is an increasingly popular framework for developing robotic applications. Its second major version, denoted as ROS~2, was designed to fill the needs of industrial use cases, including support for real-time execution. ROS 2 supports writing applications in different programming languages by providing so-called client libraries. Out of the box, ROS~2 provides client libraries for C++ and Python; other languages, such as Rust, have various levels of community support.

Despite its detailed design, early versions of ROS~2 did not deliver optimal timing predictability~\cite{2019:Casini}. The discovered problems were since fixed and researchers started developing response time analysis techniques for various configurations of ROS~2 applications~\cite{2020:Tang,2021:Blass,2022:Jiang,2024:Teper}. All these techniques assume the use of C++ language, which is officially supported by the framework and is considered mature and suitable for real-time applications.

However, even C++ support in ROS is not without problems. Teper et al.~\cite{2024:Teperc} discovered that ROS multi-threaded executor is not starvation-free, making it unsuitable for analysis with existing techniques. Another problem often associated with C++ is its complexity. Writing reliable C++ applications becomes difficult even for professionals and even more so for less experienced users. That is one of the reasons why the Rust programming language gains in popularity. Its compiler can detect many types of common C++ errors, such as race conditions, at compile time. ROS supports Rust via several community-provided libraries. Among the most popular libraries are \texttt{rclrs} from the \verb|ros2_rust| project~\cite{ros2_rust} and R2R~\cite{r2r}. The former provides an API similar to its C++ counterpart \texttt{rclcpp}, and the latter offers so-called asynchronous (async in short) API, which allows multiplexing execution of concurrent tasks in a single operating system (OS) thread. While \texttt{rclrs} received some contributions from core ROS developers, R2R was developed independently as a part of the Sequence Planner framework~\cite{2022:Dahl}.

Given the potential of the Rust language and its increasing popularity, it is essential to understand how it can be used to develop real-time robotic applications.
In this paper, we look at how the async R2R library schedules and executes the code of ROS~2 applications and compare that with the approaches used in the C++ ROS client library. Since the async Rust applications offer high flexibility in how the application is executed, we propose a particular structure suitable for real-time applications that utilize a thread prioritization and callback-to-thread mapping scheme. We evaluate this structure by measuring end-to-end latencies in a synthetic application as well as in a more complex autonomous driving case study. With the synthetic application, we empirically compare different application structures utilizing different prioritization schemes and different async Rust runtime libraries and compare them with the C++ language. Our proposed structure achieves bounded response times for time-critical tasks, which is more suitable for real-time applications than the structures appearing in R2R documentation and example code.
This opens the way for future work to either adapt existing response-time analysis techniques or design new ones to target R2R applications using our structure.


The specific contributions of the paper are:
\begin{enumerate}
    \item We analyze the execution model of the async Rust R2R library and of several async Rust runtimes and compare them with the execution model of C++ ROS applications.
    \item We propose how to structure R2R applications to be suitable for deterministic real-time operation.
    \item We demonstrate that the response times of the synthetic application match the theoretical results of a uni-processor response-time analysis. Moreover, a more complex autonomous driving case study shows that deterministic timing is maintained even in processing chains involving more than two nodes and running concurrently with other chains.
\end{enumerate}



\section{Background}

This section describes the basic concepts of ROS~2, followed by an introduction to async programming in Rust and a description of scheduling execution in Rust async runtimes. Then, we compare features supported by the official C++ client library \texttt{rclcpp} and two Rust client libraries R2R and \texttt{rclrs}.
Finally, we describe the execution model of the R2R library and compare it with the model of C++ ROS executors.

\subsection{ROS 2}

ROS~2 applications are composed of nodes, which can communicate with each other in a publish-subscribe manner. \emph{ROS node} is an organizational unit for other \textit{entities} such as timers, publishers and subscriptions. In this paper, we assume the traditional model with one ROS node per OS process, but ROS 2 also supports another model where multiple so-called composable nodes to run in a single process. A \textit{publisher} allows sending messages to the associated \emph{topic} and all \textit{subscriptions} to the same topic then receive the message. Reception of the message or expiration of a timer results in invocation of an associated application \emph{callback}. When the callbacks are executed depends on the \textit{executor} associated with them. The C++ client library \texttt{rclcpp} provides a single-threaded executor, a multi-threaded executor, and an experimental events executor. Besides publish-subscribe communication, ROS~2 applications can also communicate via services and actions, which are internally built on top of publishers and subscribers.  We do not explicitly address them in this paper, however their callback scheduling and time related aspects should not be much different from publishers and subscribers.

\begin{figure}
  \centering
  \includegraphics[scale=0.9]{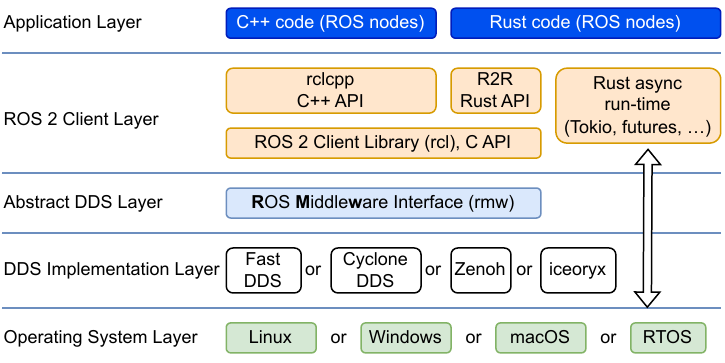}
  \caption{ROS 2 architecture for C++ and Rust R2R applications.}
  \label{fig:ros-arch}
\end{figure}

Internally, ROS 2 implementation uses a layered architecture depicted in \cref{fig:ros-arch}. The figure also shows the main differences between C++ and Rust R2R nodes, which will be detailed below.

The communication between ROS nodes is handled by the ROS middleware (RMW) library, which supports different implementations of the actual communication services. Currently, ROS~2 provides several implementations of the Data Distribution Service (DDS) standard~\cite{DDS1.4}, with Fast DDS~\cite{FastDDS} being the default, and the latest ROS~2 release Jazzy Jalisco~\cite{2024:ROS2Jazzy} adds experimental support for the Zenoh protocol~\cite{Zenoh}.

ROS client libraries can receive information about events occurring on entities, e.g., timer expirations or receptions of messages on subscriptions, via the \textit{wait set} data structure, which allows the client thread to wait for multiple entities simultaneously. After waiting, the wait set reports which entities in the set are \textit{ready}, i.e., one or more events occurred on them. The ready entities are reported in the same order in which the entities were added to the set. The process of obtaining ready entities is called \textit{sampling}.

After a subscription is ready, its received message can be obtained from RMW with an operation called \textit{take}, which is implemented by the \verb|rcl_take| function.

\subsection{Rust \& asynchronous programming}

Similarly to C++, Rust is a compiled programming language offering higher levels of abstraction than the C language but still giving the programmer detailed control over the usage of resources such as execution time and memory. An important difference from C++ is that the Rust compiler can guarantee memory safety, meaning that many classes of memory-related errors (race conditions, use-after-free) are detected and prevented at compile time.

The Rust language has support for asynchronous (\textit{async} in short) programming, which is a form of concurrency where scheduling decisions are made in the application rather than by the OS, as is the case with thread-based concurrency. To support async programming, the Rust language defines two keywords \texttt{async} and \texttt{await}, and the \texttt{Future} trait\footnote{Traits in Rust are a concept similar to interfaces or abstract classes in other languages.}, but leaves the implementation of the trait and related schedulers to independent libraries (crates in Rust terminology) called async runtimes. Popular async runtimes are Tokio\footnote{\url{https://tokio.rs/}} and \textsc{futures}\footnote{\url{https://crates.io/crates/futures}} (in small caps to avoid confusion with futures objects).

Async runtimes work with objects implementing the \texttt{Future} trait called  \textit{futures}. They represent a future execution of some code optionally producing a value. Applications can create futures in several ways: by calling a function annotated with the \texttt{async} keyword, by defining an \texttt{async} block, or by implementing the Future trait manually. Futures can be turned into \textit{async tasks} by registering them with an executor, typically by calling its \texttt{spawn} method. \textit{Executors}, which are usually provided by async runtimes and are conceptually similar to ROS C++ executors, then schedule and execute async tasks in one or more OS threads. Note that the overhead of async tasks is much lower than the overhead of OS threads. Async runtimes are known to easily handle millions of tasks. Futures can also be executed in the currently running task, without spawning a new one, by using the \texttt{await} keyword on them.


Async tasks can communicate by sending messages via async MPSC (Multi-Producer Single-Consumer) channels. Such a channel is typically implemented by a concurrent FIFO queue that allows asynchronous waiting on the receiving side until there is a message to dequeue. Technically, waiting is performed by calling a method on the receiving side that returns a future. Note that MPSC channel implementation and API can differ in different runtimes, but as long as they implement the \texttt{Future} trait, they are often compatible with other runtimes.


\subsubsection{Execution model of \textsc{futures} runtime}

We start our description of the Rust asynchronous execution model by describing the \textsc{futures} runtime. Its implementation is simple, which allows building deterministic real-time applications on top of it. It allows mapping of async tasks to OS threads, but does not control scheduling parameters of the threads in any way, allowing the application to set them as appropriate.

\paragraph{Futures local executor}

This section describes the current (v0.3.31) behavior of the \verb|LocalPool| (local in short) executor within the \textsc{futures} crate. The local executor is single-threaded and incorporates three primary data structures: 1)~an incoming tasks vector, 2)~a linked list of active tasks, and 3)~a ready queue implemented by a concurrent linked list. When a future is spawned into the executor, a new async task is allocated in heap memory and added to the incoming vector. The executor repeatedly checks the incoming vector for new tasks and moves them to the list of active tasks, which incurs additional memory allocation for each task. Subsequently, the task is enqueued into the ready queue to execute the future or set up waiting for it. The executor ensures that each task is present in the ready queue only once at any given time. To execute a ready task, the executor removes the task from the ready queue and then invokes the \verb|Future::poll| method on it. 
Whenever an active task (executing or waiting) becomes ready, it is always enqueued at the end of the ready queue.

\paragraph{Futures thread-pool executor}

The thread-pool executor executes tasks in a pool of multiple worker threads. It uses only the ready queue data structure, which is implemented using the standard library's unbounded synchronous MPSC channel, i.e., concurrent linked list. When a waiting task is woken, it is added to the end of the ready queue. This operation can involve memory allocation. Threads in the thread pool dequeue ready tasks in FIFO order in mutually exclusive way and run them. If the queue is empty, the thread waits. The process of adding a running task back to the ready queue (without waiting) differs from the local executor in that the task is not added to the end of the ready queue but continues executing. This might result in starvation of other tasks. For example, if in a thread pool with $N$ worker threads, there are $N$ tasks that always become ready during their execution, other tasks will not be executed at all.


\paragraph{Grouping futures execution}
\label{sec:joining}
The \verb|futures::join!()| macro can be used to group multiple futures and wait for the completion of the group as a whole. The effect of joining the futures is that their code will not be executed in parallel. From the point of view of the executor, the group is treated as one task; if one future gets ready, the entire group becomes ready. When the group starts executing, it will poll all ready tasks in the group. The effect of this grouping in the executor is shown later in an example with ROS in \cref{fig:join-execution-order}.

\subsubsection{Tokio.rs runtime}

Tokio is a popular multi-threaded runtime. It seems to be designed to maximize throughput rather than time determinism. Their scheduling policies are complex and difficult to understand due to the use of abstraction layers.
The scheduling policies are partially described in the documentation~\cite{Tokio:MultiThreadedDoc} and can be outlined as follows: Each worker thread has a LIFO slot, which is used for dequeuing ready tasks at most three times in a row. We believe, this improves cache locality while also avoiding starvation. Then, tasks are dequeued from the local ready queue, which can hold up to 256 tasks. If the worker cannot dequeue tasks from the local or global (shared) ready queues (they are empty), it steals half of the tasks in the other worker's local queue. The victim of the theft is chosen as the first worker, with a non-empty queue when iterating workers with a random starting position. Such a behavior is clearly unsuitable for real-time applications.
We will evaluate Tokio experimentally in \cref{sec:evaluation}.

\subsection{ROS 2 C++ executors}

Execution of callbacks in C++ ROS~2 applications is handled by ROS executors. Currently, ROS~2 includes a single-threaded executor, a multi-threaded executor, and an experimental events executor~\cite{2024:DiscourseExecutors}. These are briefly described below.

\begin{description}
    \item[Single-threaded executor] first samples the associated entities and then executes callbacks of those that are ready. Callbacks are executed in the same thread and are ordered based on their type. Timers are first, followed by subscriptions, services, and clients. Callbacks of the same type were executed based on the order of their registration~\cite{2021:Blass}, but this has changed in ROS~2 Jazzy, where the order is no longer predictable\footnote{\url{https://github.com/ros2/rclcpp/issues/2532}}.
    \item[Multi-threaded executor] executes callbacks in multiple threads. Callbacks are organized in callback groups, which can be of two types: \textit{Mutually exclusive} or \textit{Reentrant}. The executor threads access the wait set in a mutually exclusive manner; a thread that gets the access waits for the events, and once some entities are ready, their callbacks get executed by one or more threads subject to the policy of their callback group. Multi-threaded executor was found not to be starvation-free~\cite{2024:Teperc}, which is problematic for real-time applications and their analysis.
    \item[Events executor] is a recently added experimental executor that does not use wait sets but pushes events to the executor's event queue directly from DDS callbacks. The main executor thread then dequeues the events and executes the associated callbacks in a loop. Recently, a multi-threaded version of the Events executor was proposed~\cite{2024:Machowinski}. 
\end{description}

\subsection{Feature comparison of rclcpp, R2R and rclrs}
\label{sec:comparison-rclrs-r2r}

Before looking in detail at the R2R client library and its execution model, we provide a high-level comparison of features implemented in R2R and the other ROS Rust client library \texttt{rclrs}. As both are community-supported, they both lack some features implemented in the C++ client library \texttt{rclcpp}. The comparison of all three libraries is summarized in \cref{tab:rclrs-vs-r2r} and commented in more detail below.

R2R and \texttt{rclcpp} implement all communication styles implemented by ROS, but \texttt{rclrs} has only limited support for actions. Only the action message types are available for \texttt{rclrs} applications. If the application wants to use the actions implemented in another node, it has to implement all action logic and state machines itself.

All three libraries support a specific type of communication, built on top of services, that allows to work with node parameters. R2R does not yet fully support parameter ranges, which can be used to announce the permissible range of parameter values. Ranges are used by some GUI tools like \texttt{rqt} to provide ``sliders'' for changing the parameter values. Another difference in parameter handling is in how different libraries implement parameter locking to prevent concurrent accesses from the middleware and the application. \texttt{rclcpp} leaves locking up to the application, R2R uses a single lock per node, whereas \texttt{rclrs} has one lock for each parameter, potentially causing higher overhead in nodes with high number of parameters. R2R has a feature not available in other libraries, which simplifies working with parameters by using Rust's \texttt{derive} macro to automatically generate parameter handling code for fields of an arbitrary structure.

With respect to time handling, a drawback of \texttt{rclrs} is the unavailability of timers. Time-based execution of user code has to be implemented by using standard Rust means, which prevents the correct function of such nodes with ROS simulated time. However, note that simulated time is supported, but only for clocks and not for timers. Another time-related feature is support for tracing, which is being submitted to R2R as a result of this work~\cite{2025:SkoudlilThesis}. Similar functionality is missing in \texttt{rclrs}.

The supported executors in R2R are detailed in the next section. Here, we just mention that \texttt{rclrs} supports only a single-threaded executor. Execution in multiple threads can be implemented by using multiple single-threaded executors or by sending work from callbacks to other threads via standard Rust means.

Neither Rust client library supports composable and lifecycle nodes. While the latter could be implemented with little effort, the former would require a deeper investigation of ABI compatibility between Rust and C++.

\begin{table}
\begin{tabular}{|l|l|l|l|}
\hline
Feature                         & rclcpp & R2R & rclrs\\
\hline
\textbf{Communication}          &  &  & \\
Message generation              & \YES  & \YES  & \YES \\
Publishers and subscriptions    & \YES  & \YES  & \YES \\
Loaned messages (zero-copy)     & \YES  & \YES  & \YES \\
Tunable QoS settings            & \YES  & \YES  & \YES \\
Clients and services            & \YES  & \YES  & \YES \\
Actions                         & \YES  & \YES  & \WARN\footnotemark[1]\\
Dynamic type support            & \YES  & \YES  & \YES \\
\hline
\textbf{Parameters}             &  &  & \\
Parameter handling              & \YES  & \YES  & \YES \\
Parameter ranges                & \YES  & \WARN & \YES \\
Parameter locking               & none & per-node & per-parameter\\
Derived parameters\footnotemark[2] & \NO   & \YES  & \NO  \\
\hline
\textbf{Time}                   &  &  & \\
Timers                          & \YES  & \YES  & \NO  \\
Simulated time                  & \YES  & \YES  & \YES \\
Tracepoints                     & \YES  & \YESx\footnotemark[3] & \NO  \\
\hline
\textbf{Executors}              &  &  & \\
Single-threaded                 & \YES  & \YESx\footnotemark[4] & \YES \\
Multi-threaded                  & \YES  & \YESx\footnotemark[4] & \NO  \\
Asynchronous programming style  & \NO   & \YES  & \NO  \\
\hline
\textbf{Other}                  &  &  & \\
Composable nodes                & \YES  & \NO   & \NO  \\
Lifecycle nodes                 & \YES  & \NO   & \NO  \\
\hline
\end{tabular}

\medskip
\noindent Legend:\\
\begin{tabular}{ll}
  \begin{minipage}[t]{0.4\linewidth}
    \YES{}  Supported\\
    \YESx{} Supported with comment
  \end{minipage}
  &
  \begin{minipage}[t]{0.4\linewidth}
    \WARN{} Partially supported\\
    \NO{} Not supported
  \end{minipage}
\end{tabular}

\medskip
{\footnotesize
\footnotemark[1] See
\url{https://github.com/ros2-rust/ros2\_rust/issues/244},
\url{https://github.com/ros2-rust/ros2\_rust/pull/423},
\url{https://github.com/ros2-rust/ros2\_rust/pull/410}.

\footnotemark[2] Automatic generation of parameter handling code for fields of a structure.

\footnotemark[3] Implemented in pull request \url{https://github.com/sequenceplanner/r2r/pull/117},
likely to be merged soon.

\footnotemark[4] Executors are not a part of R2R, but are provided by
asynchronous runtime libraries like \textsc{futures} or Tokio. Hence, the
exact types of supported executors depend on the selected library.

  \caption{Feature comparison between rclcpp, R2R and rclrs.}
  \label{tab:rclrs-vs-r2r}
}
\end{table}

\subsection{R2R execution model}

R2R~\cite{r2r} is one of ROS~2 client libraries for Rust. Its execution model differs from C++ because callback execution is managed by a Rust async runtime rather than by R2R itself as shown in \cref{fig:ros-arch}. R2R is only responsible for ``sampling'' the events from ROS entities and pushing them to the async runtime via MPSC channels.



To sample the events, R2R uses wait sets as the official C++ executors. Sampling is performed by function \verb|Node::spin_once|. It creates a wait set with all entities in the ROS node (subscriptions, timers, clients, and services), 
and then it waits on it until one or more entities are ready. Until this point, the behavior is the same as in C++. Then, the ready entities are iterated over, received messages are taken (by calling \verb|rcl_take|) from the RMW, and all events (messages and timer expirations) are pushed to async MPSC channels associated with their entity. R2R uses bounded asynchronous channels from the \textsc{futures} crate, which are mapped 1:1 to entities. In the current implementation, the channels have a fixed capacity of 11 events. 
If the channel is full, new events are dropped, leading to unbounded response time (the dropped callback instances will never be executed). 

The registration of entity callbacks differs from C++, where methods for creating timers or subscriptions take the callback as a parameter. In R2R, the corresponding methods return the receiving end of the associated MPSC channel, and the callback is bound to it by spawning an async task created from an async block, which periodically awaits the timer expiration or the message from the channel. See \cref{list:r2r-callback} for an example. This way, if a subscription channel contains more than one message, the callback is consecutively executed for all messages available in the channel.

\begin{listing}
  \caption{Registration of a subscription callback in R2R.}%
\begin{minted}{rust}
let subscription_future = subscription.for_each(|msg| async move {
    // Callback code for msg processing here
});
executor.spawn(subscription_future);
\end{minted}
  \label{list:r2r-callback}%
\end{listing}

\begin{listing}
  \caption{R2R setup equivalent to one single-threaded executor in C++}
\begin{minted}[linenos]{rust}
local_executor.run_until_stalled();       // initialization
loop {
    node.spin_once(Duration::seconds(1)); // sampling
    local_executor.run_until_stalled();   // execution
}
\end{minted}
  \label{list:r2r-spin-loop-interleaved}
\end{listing}

The execution of the callbacks is carried on by the executors. An example of the main R2R loop with the \textsc{futures} local executor is shown in \cref{list:r2r-spin-loop-interleaved}.
The call to \verb|run_until_stalled| before the loop at line 1 is necessary if we want to avoid memory allocation in the loop. The call initializes the executor by moving the tasks from the incoming vector to the active list, which involves memory allocation. Since the \verb|spin_once| function was not called before, no tasks were sampled, and no callback code is going to be executed.

When calling the \texttt{spin\_once} function provided by R2R (line 3), it wakes up receiving tasks of all ready entity channels. Subscriptions first, then timers, etc. Note that this order differs from C++ executors. The order of waking tasks of the same entity type follows the creation order of the entities in the node.

The second call to \verb|run_until_stalled| inside the loop at line 4 executes the callbacks. Since the local executor schedules tasks in the FIFO manner, the tasks will run in the order in which they were woken up.

In such a setup, each channel will store at most one message because for each ready entity \verb|spin_once| pushes only a single message to its channel. A subsequent call to \verb|run_until_stalled| then processes all channels, leaving them empty. Therefore, the channel will never drop messages due to it being full.

\begin{figure}
    \centering
    \includegraphics[width=1\linewidth]{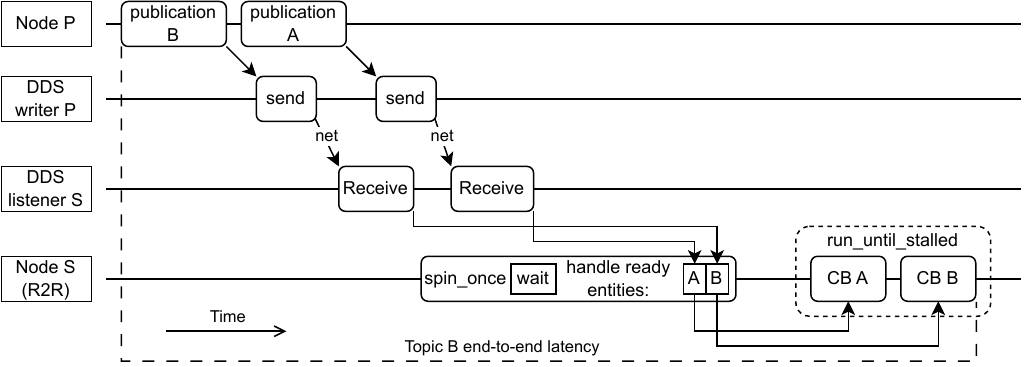}
    \caption{Example of execution of a chain of operations from publishing two messages in one node to processing it with R2R in another node. Horizontal lines represent different involved threads. The subscriptions in R2R were created in alphabetical order of topics.}
    \label{fig:gantt-local-pool-single-thread}
\end{figure}

\cref{fig:gantt-local-pool-single-thread} shows the above described sequence of operations in time. 
Besides the R2R thread running the loop from \cref{list:r2r-spin-loop-interleaved} (denoted as Node S), it shows publisher and DDS threads involved in the process.

Note that the callbacks may not always be executed in the FIFO order as described above. One such case can happen if \verb|spin_once| is called multiple times without calling \verb|run_until_stalled| in between. This is demonstrated in the example in~\cref{fig:multiple-pending-of-same}. The rightmost column shows in which order would be the callbacks called by a call to \verb|run_until_stalled| following the \verb|spin_once| calls.

\def\gapbox#1{{\setlength{\fboxrule}{0pt}\fbox{\setlength{\fboxrule}{0.5pt}#1}}}
\begin{figure}
    \centering
        \begin{tabular}{|>{\centering\arraybackslash}p{2cm}|p{2cm}|p{3.2cm}|p{5cm}|}
            \hline
           \texttt{spin\_once} call& Ready entities after sampling &  Executor ready queue state after sampling& Order of callback execution by the first call to \texttt{spin\_until\_stalled} after sampling.\\
           \hline\hline
           1 & B D & \gapbox{←\fbox{B} \fbox{D}←} & B$_1$, D\\
           \hline
           2 & A B C &  \gapbox{←\fbox{B} \fbox{D} \fbox{A} \fbox{C}←} & B$_1$, B$_2$, D, A, C\\
           \hline
        \end{tabular}
        \caption{Content of executor FIFO ready queue when two subsequent sampling calls return the same entity (B), producing two messages B$_1$ and B$_2$.}
    \label{fig:multiple-pending-of-same}
\end{figure}

A similar effect can happen if multiple futures are joined as described in Section~\ref{sec:joining}. An example of how the callbacks would be executed when grouping them via the join macro is shown in \cref{fig:join-execution-order}.

\begin{figure}
    \centering
        \begin{tabular}{|c|p{2.5cm}|p{5cm}|}
            \hline
           Variant & Ready entities (in creation order) &  Executor ready queue state after sampling with joined A, C, and E\\
           \hline\hline
           1 & A B C D E & \gapbox{←\fbox{\mbox{\fbox A\fbox C\fbox E}} \fbox B \fbox D←} \\
           \hline
           2 & B C D E&  \gapbox{←\fbox B \fbox{\fbox C\fbox E} \fbox D←} \\
           \hline
        \end{tabular}
    
    \caption{Execution order of the join group depends on the waking time of the joined tasks. In variant 1, the group is woken by waking of task A. In variant 2, task A is not ready, and the group is woken by task C.}
    \label{fig:join-execution-order}
\end{figure}






\begin{figure}
    \centering
    \includegraphics[width=0.8\linewidth]{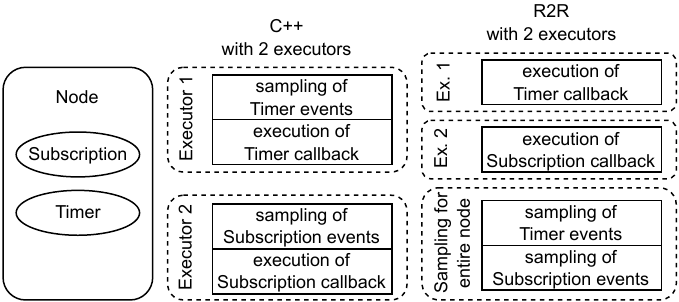}
    \caption{The sampling events of node's entities cannot be split into subsets in R2R, unlike in C++, but the callback execution can be assigned to different executors.}
    \label{fig:difference-multiple-executors}
\end{figure}

\subsection{Comparison of C++ and R2R execution models}

The main difference between C++ and R2R execution models is summarized in \cref{fig:difference-multiple-executors}. R2R does not sample events in executors because waiting on a wait set is a synchronous blocking operation, which should not be executed from an async task.

However, if the R2R application is organized as in \cref{list:r2r-spin-loop-interleaved}, the execution is equivalent to using a ROS C++ single-threaded executor with a minor difference that the priorities of subscription and timer callbacks are swapped. This means that the response-time analysis proposed in~\cite{2021:Blass} can be applied to R2R as is while taking into account the difference in priorities.

\begin{listing}
  \caption{R2R sampling loop to mimic ROS C++ multi-threaded executor.}
\begin{minted}{rust}
loop {
    node.spin_once(Duration::seconds(1));
}
\end{minted}
  \label{list:r2r-spin-only-loop}
\end{listing}

To achieve the behavior of a C++ multi-threaded executor with R2R, sampling must be performed in a dedicated thread with a loop shown in \cref{list:r2r-spin-only-loop}. The effect of different C++ callback group configurations can be accomplished as follows:

When all callbacks are executed in another dedicated thread with \textsc{futures} local executor, e.g.\ by calling just \verb|local_executor.run()|, this is an analogy to having all callbacks in the same mutually exclusive group.

When callbacks are executed by \textsc{futures} thread-pool executor or Tokio runtime, this is equivalent to using ROS C++ multi-threaded executor with each callback in its own mutually exclusive group, i.e., each callback can run concurrently with other callbacks but not with itself. It is possible to combine tasks to create a larger mutually exclusive group by using the \texttt{join} macro.

To achieve the same behavior as the reentrant callback group of ROS C++ multi-threaded executor, where the tasks can be executed concurrently even with itself, the callback code should dynamically create tasks and spawn them to the Rust multi-threaded executor (e.g. \textsc{futures} ThreadPool or Tokio).

\section{Achieving deterministic real-time operation with R2R}
\label{sec:proposal}
To configure R2R for use in real-time applications, we propose to run the application under the Linux SCHED\_FIFO scheduler and structure it as summarized by the following two rules:
\begin{itemize}
    \item The main thread should run with the highest priority and should sample events from ROS, i.e., it should call \verb|Node::spin_once| in a loop.
    \item The callbacks should be executed in lower-priority threads, each running one \textsc{futures} local executors.
\end{itemize}
An example of the proposed structure is given in \cref{list:r2r-proposed}. The priority of the main thread is set on lines 14--18, before creating ROS context and node at lines 20 and 21. A single subscriber is then created on line 23, followed by associating it with a callback implemented as an \texttt{async move} block (lines 24--26). Line 27 spawns a new OS thread with the given priority. The code executed by the thread is a closure defined at lines 5--10. It creates a new local executor (line 6), spawns to it a new async task from the passed future (line 8) and runs the executor loop (line 9).

Running the main thread with the highest priority serves two purposes. First, RMW/DDS threads, which are created internally during ROS initialization, inherit the same priority and, therefore, will never be delayed by callback execution. Second, sampling of entities for events will happen almost immediately after the event happens, without waiting for any callback to complete.

Execution of callbacks in lower-priority threads ensures that one can use schedulability analysis, which does not need to take into account ROS specifics and depends only on the scheduling policy of the OS scheduler. In \cref{sec:evaluation}, we demonstrate this by running each callback in a thread pinned to a single CPU and with priority set according to the rate-monotonic ordering and use uni-processor response-time analysis~\cite{1993:Audsley}\footnote{Application of more complex analysis techniques should be possible and is left for future work.} to predict the response times.

Note that in the current R2R implementation, where all MPSC channels have the same fixed capacity, it could happen that a channel served by a callback in a low-priority thread could overflow, leading to message losses. Therefore, for reliable operation in all situations, R2R should be extended to make channel capacity configurable, and one should use a schedulability analysis to dimension the capacity of individual channels.

\begin{listing}
  \caption{Proposed structure of a real-time R2R application}
\begin{minted}[linenos]{rust}
fn spawn_in_thread(future: impl Future, priority: ThreadPriority) {
    let thread = ThreadBuilder::default()
        .policy(RealTime(Fifo))
        .priority(priority)
        .spawn(move |_| {
            let mut local_executor = executor::LocalPool::new();
            let spawner = local_executor.spawner();
            spawner.spawn_local(future).unwrap();
            local_executor.run();
        });
}

fn main() -> Result<(), Box<dyn Error>> {
    thread_priority::unix::set_thread_priority_and_policy(
        thread_priority::thread_native_id(),
        ThreadPriority::try_from(MAIN_PRIORITY)?,
        RealTime(Fifo),
    )?;

    let ctx = r2r::Context::create()?;
    let mut node = r2r::Node::create(ctx, "example", "")?;

    let subs = node.subscribe("/topic", QosProfile::default())?;
    let future = subs.for_each(move |msg: Msg| async move {
        // process msg
    });
    spawn_in_thread(future, ThreadPriority::try_from(CALLBACK_PRIORITY)?);

    loop {
        node.spin_once(SPIN_TIMEOUT);
    }
}
\end{minted}
  \label{list:r2r-proposed}
\end{listing}

\section{Experimental evaluation}
\label{sec:evaluation}

To evaluate the proposed application structure and compare it with other alternatives, we designed a set synthetic benchmarks. These are described in \cref{sec:synthetic-benchmarks}. Evaluation on a more complex autonomous driving application follows in \cref{sec:compl-auton-driv}.

\subsection{Synthetic benchmarks}
\label{sec:synthetic-benchmarks}

\begin{figure}
    \centering
    \includegraphics[width=0.5\linewidth]{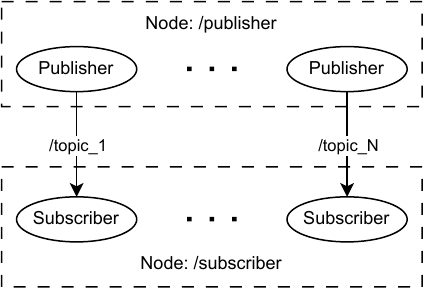}
    \caption{ROS application used for evaluation with $N=5$}
    \label{fig:test-system-nodes}
\end{figure}

To evaluate the proposed application structure and compare response times obtained with different setups, we develop a benchmark application composed of five topics (see \cref{fig:test-system-nodes}). Messages are published periodically to the topics by a publisher node running in a dedicated process. The second node subscribes to the topics and executes callbacks with a specific fixed execution time. The publication periods and callback execution times are given in \cref{tab:pub-sub-parameters} and lead to CPU utilization of 90\%.

\begin{table}
    \centering
    \begin{tabular}{|l|ccccc|}
    \hline
         \bf Topic \#&  1&  2&  3&  4& 5\\
    \hline
         \bf Publisher period [ms]&  10&  20&  50&  100& 200\\
         \bf Subscription callback execution time [ms]&  2&  4&  5&  15& 50\\
    \hline
    \end{tabular}
    \caption{Parameters of publishers and subscriptions in the benchmarking application}
    \label{tab:pub-sub-parameters}
\end{table}

We implement the subscribers in several variants using different execution strategies. The implemented variants are attached to the paper.
We run the application and trace its execution with the help of LTTng\footnote{\url{https://lttng.org/}} toolkit. For C++ variant, we reuse the tracepoints already present in the ROS libraries, for R2R variants, we add tracepoints to R2R and also to subscription callbacks. The collected traces allow the measurement of intermediate or end-to-end latencies of individual messages and callback executions in the application. Then, we calculate various statistics from them.

\subsubsection{Execution environment}

All experiments are executed on a laptop with AMD Ryzen 7 3700U CPU running Ubuntu 24.04 with the Linux kernel 6.8.0. The used ROS distribution is ROS Jazzy Jalisco. Processes that need their main thread to run with real-time priority (SCHED\_FIFO) were started via the \verb|chrt| command so that the RMW/DDS threads inherit the priority and policy from it. Other threads' priorities are set by calling \verb|pthread_setschedparam| either directly in C++ or through the Rust \verb|thread-priority| crate. 

All threads in all experiments were run on a single isolated CPU core (hyper-threading is enabled, but both threads are set as isolated; only one of them is used for experiments) by using the \verb|isolcpus| kernel command line argument. CPU affinity of the threads was set by starting the programs via the \verb|taskset| command.

\subsubsection{Publisher node implementation}

The publisher node is implemented with R2R and Tokio runtime. Publications are triggered by absolute time timers implemented via the Linux \texttt{timerfd} system call and connected to Tokio. Each timer callback is an async task that publishes a message containing a single 64-bit integer. All threads are scheduled by real-time SCHED\_FIFO scheduler. The priority of the main and RMW/DDS threads is set to 25, a single Tokio worker thread has priority of 24.

\subsubsection{Implementation variants of the subscriber node}

We implemented the following variants of the subscriber node that we compare. Below, we provide their names together with a short description.
\begin{description}
    \item[futures] R2R-based subscriber node with \textsc{futures} local executor running interleaved with the \verb|spin_once| function in the same thread, as in \cref{list:r2r-spin-loop-interleaved}.
    \item[futures-join] R2R-based subscriber node with \textsc{futures} thread-pool executor. \verb|spin_once| is running alone in the main thread. All callback tasks are joined together; the single resulting task runs in the thread pool. 
    \item[futures-rt] The variant proposed in \cref{sec:proposal}, that is R2R-subscriber node with each callback running in a separate thread with real-time priority set according to the rate-monotonic (RM) priority assignment (the smaller period, the higher priority) in a \textsc{futures} local executor. The \verb|spin_once| function is called in a loop in the main thread with the highest priority.
    \item[futures-thread-pool] R2R-based subscriber node with a \textsc{futures} thread-pool executor. \verb|spin_once| is running in the main thread, and the thread-pool running the callbacks has two threads.
    \item[futures-2-threads] R2R-based subscriber node with the \verb|spin_once| function is called in a loop in the main thread with the highest priority. The second thread executes all callbacks in the \textsc{futures} local executor.
    \item[rclcpp-rt] C++ subscriber node with each subscriber executed by its own single-threaded executor running on a dedicated thread with fixed real-time priority (RM). This is a similar scheme to the one proposed in \cite{2021:Krukiewicz-Gacek}.
    \item[rclcpp-st] C++ subscriber node with the default single-threaded executor (started by \verb|rclcpp::spin(node)|) for all callbacks.
    \item[tokio] R2R-based subscriber node using Tokio to execute callbacks. All threads in the process inherit the same priority from the main thread (Tokio's default). The \verb|spin_once| function is executed in a loop on a separate dedicated thread. Callbacks are executed by Tokio's async runtime.
    \item[tokio-rt] R2R-based subscriber node using Tokio to execute the callbacks. The \verb|spin_once| function is executed in a loop on a separate dedicated thread. The worker threads running callbacks have all the same priority lower than the spin thread.

\end{description}


In all variants mentioned above, the main thread, along with all DDS threads, has priority 21, and if callback-running threads have different priorities, they use priorities from 20 down to 16, respectively. Tokio workers in \textit{tokio-rt} have priority 20. All threads are executed with the \verb|SCHED_FIFO| scheduler.

The only exceptions are variants prefixed with \verb|nort|, in which the subscriber node with all its threads are scheduled by the default scheduler (\verb|SCHED_OTHER|).

The callback work is emulated by a loop that executes for the given time. The elapsed time is measured by calling \verb|clock_gettime()| with clock type \verb|CLOCK_THREAD_CPUTIME_ID|.

Subscriptions are created with ROS Quality of service set to keep the last 100 messages because if we kept only one message, subscriptions with a response time greater than its period would lose messages. The value 100 was chosen because it is greater than what is actually needed so as not to lose any message.



\subsubsection{Synthetic benchmark results}
\label{sec:synth-benchm-results}

\begin{figure}
    \centering
    \includegraphics{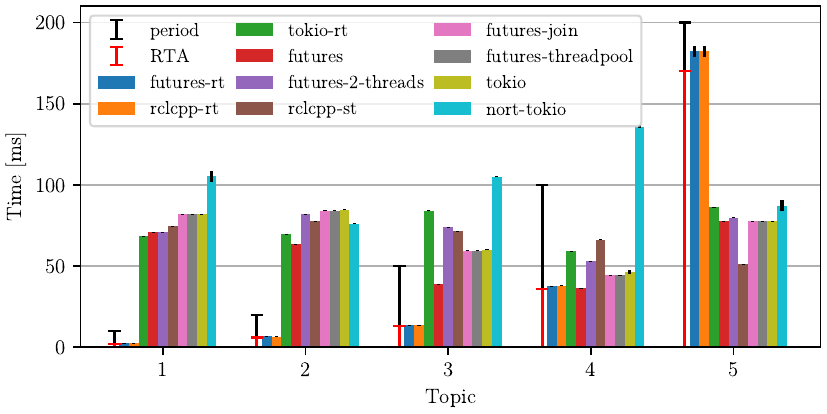}
    \caption{End-to-end latency (99$^\text{th}$ percentile) of various subscriber implementations. Black error bars represent standard deviations.}
    \label{fig:response-time-plot}
\end{figure}

We executed all implemented variants for 20 seconds, leading to the publication of 100 to 2000 messages to the respective topics. All experiments were executed 10 times, and standard deviations of different runs were calculated and reported in graphs with error bars.
The 99$^\text{th}$ percentiles of the measured end-to-end latencies (from the publication of a message to the end of callback execution) are reported in \cref{fig:response-time-plot}. The graph also shows the publication periods and the expected worst-case response time calculated with a classical mono-processor response-time analysis (RTA)~\cite{1993:Audsley}.

As can be seen, our proposed \textit{futures-rt} achieves the same results as C++ \textit{rclcpp-rt}, and the results of both variants match the theoretical worst-case response time calculated by the RTA. Only topic 5, running in the lowest-priority thread, exhibits higher latencies than expected, which is caused by overheads of the sampling and RMW/DDS threads, which are not considered by our RTA\footnote{It should be possible to model the overheads as separate tasks or associate them to existing tasks, however, it would require a more detailed execution time analysis than what we do in this paper.}. Note that \textit{futures-rt} and \textit{rclcpp-rt} are the only variants, where all callbacks manage to complete before the deadlines given by the publication periods.

All other implemented variants fail to meet shorter deadlines for topics 1--3. This is expected for the default configurations of Rust async runtimes, which are not able to take into account the timing requirements of individual callbacks. This is especially visible for the nort-tokio variant, which has the highest latencies for three of five topics.

\begin{figure}
    \centering
    \includegraphics{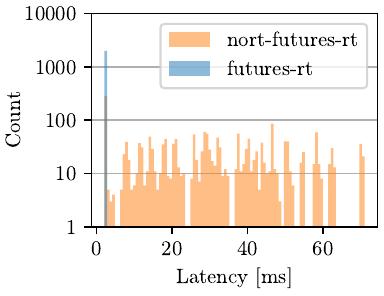}
    \includegraphics{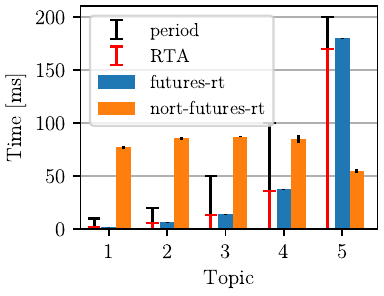}
    \caption{Comparison of end-to-end latency with and without setting the real-time priority of RMW/DDS threads. Left: Histogram of latencies of the highest-rate topic 1. Right: 99th percentile of latencies of all topics.}
    \label{fig:rmw-prio}
\end{figure}

To investigate the effect of RMW/DDS thread priority, we compare the end-to-end latencies of our proposed futures-rt with and without giving the RMW/DDS threads the highest priority. The results in \cref{fig:rmw-prio} show that without setting the priority of RMW/DDS threads to be higher than the priorities of the callback threads, the callbacks fail to complete before the deadlines given by the publication periods.

\subsection{Complex autonomous driving application}
\label{sec:compl-auton-driv}

To evaluate the suitability of R2R for implementing more complex real-time applications, we used it to implement a simplified version of the Automated Lane Keeping System (ALKS) -- a system that can automatically drive the car on highways with speed up to 130\,km/h \cite{UN-R157am4}. ALKS is the first autonomous driving system legally allowed in Europe that reaches SAE Level 3 of automation~\cite{SAE_J3016_202104}, meaning that driver supervision is not required.

\begin{figure}
  \centering
  \includegraphics[width=\linewidth]{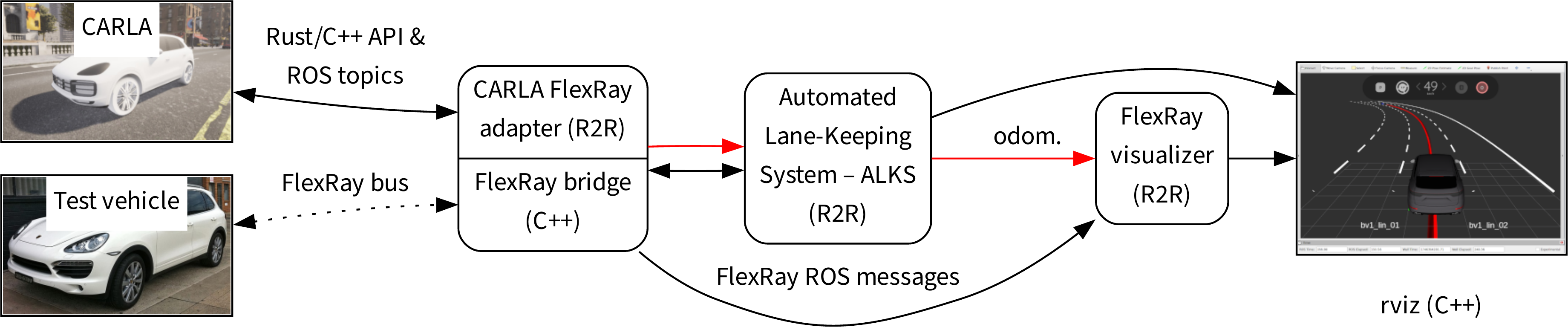}
  \caption{Architecture of the complex application when using the CARLA simulator or a real car. Red arrows represent the odometry chain evaluated below.}
  \label{fig:alks-arch}
\end{figure}

Our ALKS implementation can drive a real Porsche Cayenne car by connecting to its FlexRay buses and communicating with the onboard control units. However, in this paper, we evaluate the version running against the simulated vehicle in the CARLA simulator~\cite{dosovitskiy_carla_2017}. The reason is that this version contains three ROS nodes written using R2R, whereas the real car version has only two R2R-based nodes. The high-level architecture of the application is depicted in \cref{fig:alks-arch}.

The first R2R node is the \emph{CARLA FlexRay adapter}, which communicates with the CARLA simulator via its API (by using the \texttt{carla-rust}\footnote{\url{https://github.com/jerry73204/carla-rust}} bindings) and ROS topics\footnote{We need to combine ROS topics with CARLA API because CARLA 0.10.0 does not publish all needed information via ROS topics.}. It publishes the information from the simulator to ROS in the same format as used when interfacing the real car via the C++-based \emph{FlexRay bridge} node. The published information consists of eleven topics, which include quantities like speed, steering wheel angle, GPS position, odometry, Inertial Measurement Unit (IMU), information about other visible vehicles, and detected road lines. Most of the quantities are published with a frequency of 50\,Hz, and information about vehicles and lines with a frequency of 25\,Hz. Producing road line messages involves the execution of a non-trivial curve-fitting algorithm for the conversion of simulated lines to their parametric spline representation used on the FlexRay bus. In the opposite direction, these nodes convert the control commands for acceleration/braking and steering received via ROS to the simulator requests or the FlexRay messages. 

The implementation of ALKS resides in the ALKS ROS node. The information received via eleven topics from the FlexRay is processed in the respective callbacks, where it is converted with simple calculations to the internal representation and stored in memory for later use. The main computation happens in a callback of a 50\,Hz timer running in a separate thread, where all logic and PID controllers are calculated. Besides that, the computationally more demanding Model-Predictive Controller (MPC) that calculates optimal trajectory is invoked from the same callback every 300\,ms. At the end of every callback invocation the node publishes control commands for the vehicle as well as debugging and visualization topics. Topics related to the current trajectory are published whenever a new one is calculated.

The last R2R-based node is the \emph{FlexRay visualizer}. It receives information from the real or simulated FlexRay bus and converts it to rviz markers, i.e., ROS messages, which can be visualized in 3D by the ROS \texttt{rviz} tool. It also receives preprocessed odometry information from ALKS, which is used for the visualization.

The overview of threads in the nodes and their SCHED\_FIFO priorities is given in \cref{tab:alks-threads}. Only the ALKS node runs callbacks in more than one thread.

\begin{table}
  \begin{center}
    \begin{tabular}{lllrl}
      \hline
      Node       & Thread(s)                                  & Callbacks   & Priority & WCET\footnotemark[1] \\
      \hline
      FlexRay    & DDS                                        & ---         & 20       & ---                       \\
      adapter    & loop \{ spin\_once \}                      & ---         & 10       & ---                       \\
                 & run                                        & all         & 10       & 0.3\,ms$^\dagger$ \\
                 & CARLA client threads                       & ---         & 10       & ---                       \\
      \hline
      ALKS       & DDS                                        & ---         & 20       & ---                       \\
                 & loop \{ spin\_once; run\_until\_stalled \} & subscribers & 10       & 0.13\,ms$^*$ \\
                 & run                                        & timer       & 5        & 9.8\,ms                    \\
      \hline
      FlexRay    & DDS                                        & ---         & 20       & ---                       \\
      visualizer & loop \{ spin\_once; run\_until\_stalled \} & all         & 10       & 0.26\,ms$^*$ \\
      \hline
      \makebox[0em][l]{
      \begin{minipage}[t]{\linewidth}
        \smallskip
        {\footnotesize
          \footnotemark[1] 99$^{\text{th}}$ percentile\\
          $^\dagger$ WCET of the odometry callback only. \\
          $^*$ WCET of run\_until\_stalled, which can invoke one or more callbacks.
        }
      \end{minipage}
      }
    \end{tabular}
  \end{center}

  \caption{Overview of threads in R2R nodes of the ALKS application.
    All threads are scheduled using SCHED\_FIFO and those running callbacks use \texttt{run*} methods of \textsc{futures} local executor.}
  \label{tab:alks-threads}
\end{table}

\subsubsection{Latency evaluation}
\label{sec:latency-evaluation}

To show that R2R is suitable for implementing practical real-time applications, we run the above-described ALKS application and trace it with LTTng. As opposed to the synthetic benchmark, here we did not restrict the threads to run on a single CPU core. From the traces, we obtain end-to-end latencies of the \emph{odometry} chain (marked with red arrows in \cref{fig:alks-arch}). This chain goes through all three R2R-based nodes in the application. Specifically, it starts at the reception of the IMU message from CARLA in the FlexRay adapter. In the associated callback, we retrieve other information from CARLA using its API and publish the obtained data to the corresponding ROS topic. The ALKS node then receives (among others) the odometry message, converts it from relative to absolute values, and stores it for later use. It also publishes the converted absolute values for use in the FlexRay visualizer. Flexray visualizer just stores the received odometry values in memory for use in other callbacks.

\begin{figure}
  \begin{subfigure}[t]{0.48\textwidth}
    \centering
    \includegraphics{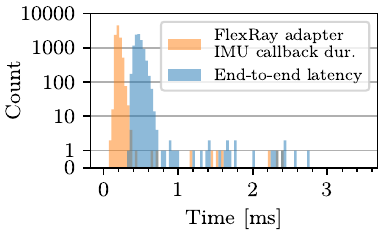}
    \caption{Histogram of end-to-end latency of the odometry chain in comparison with IMU callback duration in FlexRay adapter, which starts the chain.}
    \label{fig:alks-e2e}
  \end{subfigure}
  \hfill
  \begin{subfigure}[t]{0.48\textwidth}
    \centering
    \includegraphics{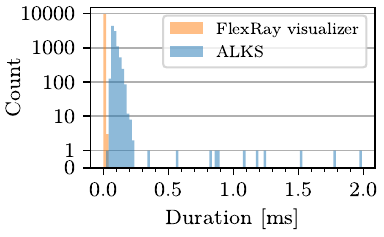}
    \caption{Histogram of odometry callback durations in ALKS and FlexRay visualizer nodes.}
    \label{fig:alks-odo-callbacks}
  \end{subfigure}

  \medskip
  \begin{subfigure}[t]{0.48\textwidth}
    \centering
    \includegraphics{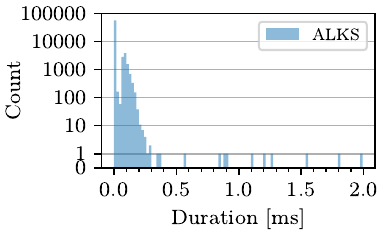}
    \caption{Histogram of \texttt{run\_until\_stalled} execution times in the subscriber thread of the ALKS node.}
    \label{fig:alks-all-subscribers}
  \end{subfigure}
  \hfill
  \begin{subfigure}[t]{0.48\textwidth}
    \centering
    \includegraphics{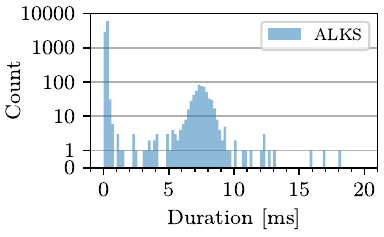}
    \caption{Histogram of timer callback duration in the ALKS node. The second peak around 8\,ms corresponds to MPC solver invocation every 15$^\text{th}$ iteration.}
    \label{fig:alks-timer}
  \end{subfigure}
  \caption{Histograms of obtained from traces of the ALKS application.}
  \label{fig:alks-histograms}
\end{figure}

\Cref{fig:alks-histograms} shows histograms obtained from a trace recorded during approximately 3 minutes of execution of the ALKS application. We removed few initial samples from the trace to filter out outliers from the warm-up phase.

The histogram of measured odometry chain end-to-end latencies is shown in \cref{fig:alks-e2e}. In about 99\% of cases, the latency is below 0.8\,ms. The same figure also shows execution time (duration) of the first callback in the chain (IMU). As this callback communicates with the CARLA simulator over the network, its duration and jitter is dominated by this communication latency and this propagates further down the chain. \Cref{fig:alks-odo-callbacks} shows execution time histograms of other callbacks in the chain. In the ALKS node, the majority of callback executions exhibits execution time jitter of about 0.2\,ms, which is caused by waiting for mutexes protecting data shared with the timer callback running in another thread. A few outliers between 0.3 and 2\,ms can be caused by the fact that the experiment was executed on a machine connected to the network and running other development tools, which might cause some interference. The jitter of the callback in the FlexRay visualizer node is very low, as the node is single-threaded and does not need to use mutexes in the callbacks.

To put this into the context, \cref{fig:alks-all-subscribers} shows histogram of \texttt{run\_until\_stalled} call execution times in the ALKS node, whose 99$^\text{th}$ WCET percentile is mentioned in \cref{tab:alks-threads}. This function calls the odometry callback as well as other subscriber callbacks in the node. Note that due to the higher number of its invocations when compared to the odometry callback, the 99$^\text{th}$ WCET percentile is smaller (0.13\,ms) than the one of only the odometry callback (0.16\,ms).

\Cref{fig:alks-timer} shows the execution time histogram of the timer callback in the ALKS node. The left peak represents the cases where the MPC planner was not invoked, whereas the peak around 8\,ms includes the cases with planner invocations. Note that the end-to-end latency of the odometry chain is influenced by the timer callback only thourgh short critical sections for accessing shared data. Long execution time of the MPC planner has no effect on the end-to-end latency of the odometry chain. This demonstrates that R2R with the \textsc{futures} runtime and the structure proposed in this paper is capable of achieving deterministic end-to-end latencies of the chain of callbacks executing in different nodes.

\section{Discussion}
\label{sec:discussion}

Designing a real-time ROS application with R2R is not much different from designing a C++ application with \texttt{rclcpp}. One needs to carefully plan mapping of callbacks to OS threads and assign thread scheduling parameters according to the timing requirements. An advantage of the R2R approach is that sampling of ROS entities can be run in a thread independent from threads running application callbacks, decreasing time between samples to almost zero. This allows analyzing callback schedulability independently of ROS sampling.

An important difference between C++ and Rust ROS applications is how the application should structure shared data. While C++ does not impose many restrictions on the structure, Rust requires explicit association of data with their synchronization primitives. Structuring the data inappropriately causes problems with Rust borrow checker as well as with unwanted blocking of unrelated threads, negatively influencing the timing.

An advantage of using the asynchronous programming style is that it can simplify application code, especially if combining data from messages from multiple topics. In such a case, the Rust compiler automatically constructs the state machines, which would need to be implemented manually when using synchronous programming style. This simplified the design of our CARLA FlexRay adapter.

Another benefit of asynchronous programming is that applications can easily integrate reactions to events from different event sources. For example, ROS can be one event source and other ROS-unaware 3rd party libraries can provide other sources. Depending on the chosen asynchronous runtime, this can take advantage of using efficient event demultiplexing system calls like \texttt{epoll} or \texttt{io\_uring}.



\section{Related work}

The study of real-time scheduling in ROS 2 began with Casini et al.~\cite{2019:Casini}, who modeled and analyzed mainline ROS 2. Their work led to corrections that enabled the development of formal schedulability analyses. Building on these foundations, researchers proposed schedulability analyses for the C++ single-threaded executor~\cite{2021:Blass} or designed and analyzed a new custom executor~\cite{2021:Choia}. Subsequent investigations into ROS multi-threaded executors~\cite{2022:Jiang,2023:Sobhani,2024:Teperc} not only advanced schedulability analyses but also uncovered implementation flaws in the executor design. Beyond the scheduling in the executors, researchers have approached the latencies of DDS communication middleware from both experimental~\cite{2021:Kronauer} and analytical~\cite{2023:Sciangula} perspectives, providing a more comprehensive understanding of real-time performance in ROS 2 systems.

\section{Conclusion}

In this paper, we analyzed the execution model of ROS~2 applications written in asynchronous Rust with the R2R ROS client library. The execution model is flexible, allowing the application designers to choose from many variants and asynchronous runtimes. We experimentally evaluated many of the variants and end-to-end latencies achieved with them. We proposed the \textit{futures-rt} structure, which decouples sampling and callback execution to different threads, prioritizes sampling and DDS threads, and runs callbacks in dedicated lower-priority threads. This structure achieves the best results among all evaluated options and meets all deadlines. The measured latencies match well the theoretical response times calculated with the classical response-time analysis. Practical applicability of the proposed approach was demonstrated on an autonomous driving application implementing the Automated Lane Keeping System.




\bibliography{ros}

\end{document}